\documentclass[a4paper,11pt]{article}

\usepackage{graphicx}
\usepackage{amsmath}
\usepackage{amssymb}
\usepackage{amsfonts}
\usepackage{amscd}
\usepackage{feynmp} 
\usepackage{mathrsfs}
\usepackage{psfrag}
\usepackage{subfigure}
\usepackage{a4wide}

\newcommand{\CenterObject}[1]{\vcenter{\hbox{#1}}} 
\newcommand{\braket}[1]{\ensuremath{\left\langle #1 \right\rangle}}

\newcommand{\EFT}[2]{\ensuremath{\stackrel{\left(#1\right)}{#2}}}

\newcommand{\SU}{\ensuremath{\mathrm{SU}}}

\newcommand{\irrep}[1]{\bf #1}
\newcommand{\tr}{\ensuremath{\mathrm{tr}}}
\newcommand{\GeV}{\ensuremath{\mathrm{GeV}}}
\newcommand{\eV}{\ensuremath{\mathrm{eV}}}
\newcommand{\diag}{\ensuremath{\mathrm{diag}}}
\newcommand{\Ord}[1]{\ensuremath{\mathcal{O}(#1)}}

\newcommand{\CiteSeeSaw}{\cite{Minkowski:1977sc,SeeSawmore}}

\newcommand{\CiteMuTauSymm}{\cite{MuTauSymm}}

\newcommand{\CiteMuTauModels}{\cite{MuTauModels}}

\newcommand{\CiteDSBarr}{\cite{DSBarr}}

\newcommand{\CiteDSSmirnovmore}{\cite{Smirnov:1993af,DSmore}}

\newcommand{\CiteLR}{\cite{Pati:1974yy,LRmore}}

\newcommand{\CiteQLC}{\cite{Smirnov:2004ju,Raidal:2004iw,Minakata:2004xt,QLCmore}}

\newcommand{\CiteESix}{\cite{ESix}}

\newcommand{\CiteSOten}{\cite{SOten}}
\newcommand{\CiteExtGauge}{\cite{Pati:1974yy,Georgi:1974sy,SOten,ESix}}

\newcommand{\CiteLopside}{\cite{Lopside}}

\newcommand{\CiteBoundOnMN}{\cite{BoundOnM1}}

\newcommand{\CiteBiMax}{\cite{BiMax}}

\newcommand{\CiteDSorig}{\cite{DSorig}}

\begin{document}

\begin{fmffile}{dsFMF}

\begin{titlepage}

\ \vspace*{-15mm}
\begin{flushright}
TUM-HEP-586/05
\end{flushright}
\vspace*{5mm}

\begin{center}
{\Huge\sffamily\bfseries 
Screening of Dirac flavor structure in the seesaw and neutrino mixing\\[2mm]
}

\vspace{10mm}
{\large
Manfred Lindner\footnote{E-mail: \texttt{lindner@ph.tum.de}}$^{(a)}$,
Michael A. Schmidt\footnote{E-mail: \texttt{mschmidt@ph.tum.de}}$^{(a)}$,
and Alexei Yu. Smirnov\footnote{E-mail:
\texttt{smirnov@ictp.trieste.it}}$^{(a),(b),(c)}$}
\\[5mm]
{\small\textit{$^{(a)}$
Physik-Department T30,
Technische Universit\"{a}t M\"{u}nchen,\\
James-Franck-Stra{\ss}e,
85748 Garching, Germany
}}
\\[3mm]
{\small\textit{$^{(b)}$
The Abdus Salam International Centre for Theoretical Physics, I-34100 Trieste,
Italy
}}
\\[3mm]
{\small\textit{$^{(c)}$
Institute for Nuclear Research, Russian Academy of Science, Moscow, Russia
}}
\end{center}
\vspace*{1.0cm}

\begin{abstract}
\noindent
We consider the mechanism of {\it screening} of the Dirac flavor structure  
in the context of the double seesaw mechanism. As a consequence of  
screening, the structure of the light neutrino mass matrix, $m_\nu$, is 
determined essentially by the structure of the (Majorana) mass matrix, 
$M_S$, of new super-heavy (Planck scale) neutral fermions $S$. 
We calculate  effects of the renormalization group running in order to
investigate the stability of the screening mechanism with respect to 
radiative corrections. We find that screening is stable in the
supersymmetric case, whereas in the standard model it is unstable for 
certain structures of $M_S$. The screening mechanism allows us 
to reconcile the (approximate) quark-lepton symmetry and the strong 
difference of the mixing patterns in the quark and lepton sectors. 
It opens new possibilities to explain a quasi-degenerate neutrino mass 
spectrum, special ``neutrino'' symmetries and quark-lepton complementarity. 
Screening can emerge from certain flavor symmetries or Grand Unification. 

\end{abstract}

\end{titlepage}

\setcounter{footnote}{0}

\section{Introduction}

Quarks and leptons show an apparent correspondence which suggests
their common origin~\cite{Pati:1974yy}. The observation is based on the pattern of
quantum numbers and the fact that both quarks and leptons come in 
three fermionic families (generations). This allows quarks and 
leptons to be embedded into unique multiplets associated 
with extended gauge symmetries~\CiteExtGauge.     
In spite of lack of further  experimental confirmations, the quark-lepton
symmetry and unification~\CiteExtGauge\  
are still the most appealing concepts in physics beyond the Standard Model (SM).

Tiny neutrino masses emerge  naturally from the seesaw 
mechanism~\CiteSeeSaw, which implies the existence of 
right handed neutrino fields  $\nu_R$
and new  high energy scale. The seesaw scale turns out to be numerically 
rather close to the scale of Grand Unification (GU), which 
emerges from the unification of gauge couplings. It seems as 
if both gauge and fermion unification point in the same direction 
of certain Grand Unified Theory, like SO(10) GUTs~\CiteSOten. 

On the basis of quark-lepton unification (symmetry) 
one would expect similarities between the mass 
and mixing patterns of quarks and leptons. However, 
observations~\cite{Eidelman:2004wy,Strumia:2005tc} do not support this expectation. Indeed, 

\begin{itemize}

\item 
Neutrino masses and mixings differ strongly from those in
the quark sector. The hierarchy of neutrino masses, if
exists, is weaker than the quark mass hierarchy. 
The 2-3 leptonic mixing is maximal or close to maximal, 
whereas the corresponding quark mixing is very small. 
The 1-2 leptonic mixing is large but not maximal and it seems 
smaller than the 2-3 leptonic mixing. In contrast, the 1-2 quark
mixing - the largest quark mixing - is small. The only common 
feature is that the 1-3 mixings are small in both sectors. 

\item 
Data seem to indicate a particular ``neutrino symmetry''~\CiteMuTauSymm\ which 
does not show up in other sectors of the theory. This includes maximal or 
nearly maximal 2-3 mixing and relatively small 1-3 mixing. If the 
neutrino mass spectrum would turn out to be quasi-degenerate\footnote{This
would be necessary if the evidence for neutrino-less double beta
decay~\cite{Kleingrothaus:2004wj} would be confirmed.},
this could imply certain symmetry too. However, it is very difficult 
to extend the suggested symmetries not only onto the quarks but also 
to charged leptons and models ({\it e.g.}~\CiteMuTauModels) which realize such neutrino symmetries 
are quite involved. 

\item
In the context of the seesaw mechanism, the observed values of neutrino masses 
require the masses of the right-handed (RH) neutrinos to be $M_N  \lesssim  10^{14}$~GeV, 
{\it i.e.} two orders of magnitude below the Grand Unification scale, $M_{GU}$. 
Furthermore, the data implies a certain flavor structure and a hierarchy of the 
RH neutrino mass matrix which may indicate that the scale of RH neutrino 
masses emerges as a combination of the GU-scale and some other scale close 
to the Planck mass $M_{Pl}$: 
\begin{equation}
  M_{N} \sim \frac{M_{GU}^2}{M_{Pl}} \; . 
  \label{eq:scale}
\end{equation}

\item
The experimental values of the quark and lepton
mixing angles between the first and second generations appear to sum up to the maximal 
mixing angle:  
\begin{equation}
  \theta_{12} + \theta_C~   \approx \frac{\pi}{4},  
  \label{eq:qlc}
\end{equation}
with the leptonic mixing angle $\theta_{12}$ and the Cabibbo angle $\theta_C$. 
The interpretation of such a quark-lepton complementarity 
relation~\CiteQLC\ 
is rather controversial. If it is not a numerical accident,
relation~(\ref{eq:qlc}) implies some quark-lepton connection which is 
not present in ordinary GUTs. It may require some additional 
structure in the leptonic sector which produces maximal mixing and therefore 
has certain symmetry.

\end{itemize}
All these observations can be considered as hints 
against  a simple unification of quarks and leptons. 

There exist various attempts to reconcile the possible quark-lepton
symmetry (unification) and the strong difference of patterns of masses and 
mixing in two sectors.  The main problem  is that the strong hierarchy 
of eigenvalues and small mixings of the quark mass matrices imply a
certain hierarchy in the Dirac type Yukawa couplings which propagates 
due to the symmetry/unification to the lepton sector (see, however~\CiteLopside). 
One possible solution  is that dominant contribution to the neutrino 
masses, {\it e.g.}, from the Yukawa coupling with a Higgs triplet, has no 
analogy in the quark sector. However, even in this  case, it seems 
rather unnatural that some Yukawa interactions have a particular symmetry 
which is not realized for other Yukawa interactions. In the context of 
seesaw mechanism the observed features of neutrino masses and mixing can 
be related to a particular structure of the Majorana mass matrix of the 
RH neutrinos~\CiteDSSmirnovmore. 

In this paper in order to reconcile the observed pattern of the neutrino masses and 
mixings and quark-lepton symmetry, we elaborate on the 
double seesaw mechanism~\CiteDSorig.  
Specifically, we will discuss a version where the flavor structures 
of the Dirac type Yukawa couplings cancel 
completely~\cite{Smirnov:1993af,Smirnov:2004hs}, 
in the light neutrino mass matrix. 
The properties of the 
neutrino mass matrix are entirely  determined by physics 
(in particular, symmetries) at energies above the Grand unification scale.  
We will call this ``screening of the Dirac 
flavor structure'' or simply ``screening'' 
and we will show it may lead to a natural solution of the above mentioned 
problems.

The paper is organized as follows. In sec.~\ref{sec:mechanism} we describe 
the screening mechanism. We consider the renormalization group
effects 
and the stability of the mechanism with respect to radiative corrections in 
sec.~\ref{sec:RG}. Various applications of the mechanism are studied in 
sec.~\ref{sec:applications}. In particular, we consider a possibility of 
the degenerate neutrino mass spectrum, existence of  particular neutrino
symmetries, and the quark-lepton complementarity. 
In sec.~\ref{sec:GUT} we discuss how the screening mechanism can be 
realized in GUT and models with family symmetries. Conclusions are presented 
in sec.~\ref{sec:conclusions}.

\section{The flavor screening mechanism\label{sec:mechanism}}

In addition to the SM particles we introduce the three right-handed
neutrinos, $\nu_{iR} \equiv N_i^c$, ($i = 1,3,3$),  three left-handed singlets $S_i$ and  
Higgs boson singlet $\sigma$. The leptonic interactions are assumed 
to have the following features:   
\begin{enumerate}

\item 
The SM leptonic doublets $L_i$ have Yukawa couplings $Y_{\nu}$ with $N \equiv (\nu_R)^c$; 

\item 
The Majorana mass terms of the right-handed neutrinos, $N$ are forbidden or strongly suppressed;  

\item 
The right-handed neutrinos $N_i$ have Yukawa couplings $Y_N$ with singlets 
$S_i$ and a Higgs boson $\sigma$; 

\item 
The singlets $S_i$ have Majorana mass terms, thus violating the lepton number. 
\end{enumerate}
Under these assumptions the following Yukawa interactions and mass terms
appear in the Lagrangian 
\begin{equation}
    \label{eq:LeptonicLagrangian}
   -  \mathscr{L}=  L^T  Y_\nu N \phi 
    + S^T Y_N  N \sigma 
    +\frac{1}{2} S^T M_S S  +  L^T  Y_e l^c \phi_d  + h.c. \; ,
\end{equation}
(the flavor indices are omitted here). The last term in 
eq.~\ref{eq:LeptonicLagrangian} is the Yukawa coupling which generates 
masses for the charge leptons;  in the  Standard model: 
$\phi_d = \tilde{\phi} \equiv i\sigma_2 \phi^*$. 
After the Higgs bosons develop the non-zero VEVs, 
$\braket{\phi}$ and $\braket{\sigma}$, the neutrino mass matrix is 
generated in the basis $(\nu, N, S)$
\begin{equation}
 \label{eq:matrix}   
 \mathcal{M} =  \left(
      \begin{array}{ccc}
        0 & Y_\nu \braket{\phi} & 0\\
        Y_\nu^T \braket{\phi} & 0 & Y_N^T \braket{\sigma}\\
        0 & Y_N \braket{\sigma} & M_S\\
      \end{array}
    \right)\; .
\end{equation}
For $\braket{\phi} \ll \braket{\sigma} \ll M_S$ it realizes the double 
(or cascade) seesaw mechanism~\CiteDSorig. 
(The case of singular matrices $M_S$ will be considered separately in sec.~\ref{sec:singular}). 
Block diagonalization of the matrix~(\ref{eq:matrix}) leads to the 
mass matrix of light neutrinos 
\begin{equation}
\label{eq:DoubleSeeSaw}
  m_\nu^0=\left[\frac{\braket{\phi}}{\braket{\sigma}}\right]^2Y_\nu 
\left(Y_N\right)^{-1}M_S \left(Y_N^T\right)^{-1} Y_\nu^T\; .
\end{equation}

This matrix has the remarkable feature that the
Dirac type Yukawa coupling matrices cancel 
provided that  $Y_\nu$  and $Y_N$ are proportional,
{\it i.e.},  if 
\begin{equation}
 Y_\nu = c \cdot Y_N\; ,
\label{eq:relation}
\end{equation}
with  constant $c$ being  typically of the order unity. (Its  deviation from unity can be 
{\it  e.g.} due to the renormalization group effects.) 
As a result of the cancellation  the mass matrix of light neutrinos becomes
\begin{equation}
\label{eq:screening}
  m_\nu = c^2 \left[\frac{\braket{\phi}}{\braket{\sigma}}\right]^2 M_S \; ,
\end{equation}
where the neutrino mixings emerge entirely from the flavor structure in 
$M_S$~\cite{Smirnov:1993af}. The flavor structure 
of the Dirac type Yukawa couplings is completely eliminated, while it still 
determines all the features of the quark  masses and 
mixings. We will call such a cancellation  the {\it screening} of the  Dirac flavor structure 
in the lepton mixing or briefly, screening. 
Screening is  a consequence of the double seesaw 
(\ref{eq:matrix}) and the proportionality~(\ref{eq:relation}). 
These features can emerge from certain flavor symmetries or Grand 
Unification, as it will be discussed in sec.~\ref{sec:GUT}.  
The scale of neutrino masses is determined by the scales of the double seesaw. 
Taking  
\begin{equation}  
\label{eq:scales}  
 \braket{\phi} = v_{EW},~~ \braket{\sigma} \sim M_{GU}  \sim 10^{16}~ {\rm GeV},~~ 
 M_S \sim M_{Pl}  
\end{equation}  
one finds from (\ref{eq:screening})
\begin{equation}
\label{eq:scmass}
  m_\nu \sim  \left[\frac{v_{EW}}{M_{GU}}\right]^2 M_{Pl} \; ,
\end{equation}
which lies in the phenomenologically required range, while the flavor 
structure of the neutrino mass matrix, and consequently the mass ratios and
mixings are determined by the Planck scale physics~\footnote{In the paper~\cite{Vives:2005ze}  
the double seesaw matrix~(\ref{eq:matrix}) has been considered 
with only one new singlet $S$ (so that the  $Y_N$ is 
the 3 component column).  However, such a matrix can not 
reproduce the required mass spectrum of light neutrinos for 
any structure of $Y_N$. The complete mass 
spectrum consists of three Majorana neutrinos with masses 
(in our notations) $\sim M_S$, $\sim \braket{\sigma}^2/M_S$, 
$\sim   \braket{\phi}^2M_S/\braket{\sigma}^2$,
and two pseudo-Dirac neutrinos with masses 
at the electroweak scale: $\sim\braket{\phi}$. There is only one light neutrino. 
A possibility to get a correct spectrum in the model~\cite{Vives:2005ze}
is to introduce the direct non-negligible contribution to 
the RH neutrino masses (2-2 block) of the form $Y^T_{\nu} M' Y_{\nu}$, where $M'$ is 
some $3 \times 3$ matrix to be tuned to fit  the data.}.

The Majorana mass matrix of the right handed neutrinos, $M_N$, generated 
after the first seesaw equals  
\begin{equation}
\label{eq:RHmasses}   
M_N = -\braket{\sigma}^2 Y_{N}^T M_{S}^{-1} Y_N.   
\end{equation}
For the mass scales given in eq.~(\ref{eq:scales}) we obtain
$M_N \sim  M_{GU}^2/M_{Pl} \leq 10^{14}$~GeV which 
reproduce the relation (\ref{eq:scale}) required by phenomenology. 

An attractive  scenario could be  that some symmetry  in the sector 
of  singlets $S$ exists. This symmetry leads to a particular and simple 
structure of $M_S$ at high scales.  
The symmetry is broken at lower scales and incomplete 
screening, if exists,  can be related to this breaking. 
In fact,  phenomenology may require deviations from complete screening. 
Perturbations might {\it e.g.}  explain 
the observed mass splitting of the light neutrinos in the case of a degenerate 
spectrum of the singlets $S_f$. 
In this connection one can explore the 
following origins of perturbations (breaking) of the structure (\ref{eq:matrix},\ref{eq:relation}):
\begin{itemize}
\item 
Nonzero elements in the 1-1, 1-3 and  2-2 blocks of matrix in  eq.~(\ref{eq:matrix}):   
These blocks have different gauge properties and the origin of
these perturbations can be quite different.
\item 
Mismatch between $Y_{\nu}$ and $Y_N$ destroying cancellation.
\item
Perturbations in $M_S$. They may follow from the Planck scale physics. 
If the eigenvalues of $M_S$ have  some hierarchy then 
one needs to take into account the renormalization effects due to 
the Yukawa interactions of $S$, $N$ and $\sigma$ which influence the structure of 
matrix $M_S$.  

\end{itemize}
A structure of the mass matrix for 
$(\nu, N, S)$ with non-zero 1-3 block  
(the direct $\nu S$-terms) has been considered in 
ref.~\CiteDSBarr. 
Under certain conditions the matrix leads to the linear 
dependence (instead of quadratic  or no dependence in our case) 
of  the light neutrino mass matrix on the Dirac flavor structure. 
In a sense this realizes the half-screening  effect.

Notice that one can introduce the Majorana mass matrix of the RH neutrinos
in the form~\eqref{eq:RHmasses} immediately without invoking the double seesaw mechanism.
This is done in \cite{Stech:2003sb} where form 
$M_N = - \braket{\sigma}^2 Y_N^T Y_N  +\delta M_N$ 
with  $\delta M_N$ being the correction matrix has been
{\it postulated}.
In contrast, we propose a model for the  structure~\eqref{eq:RHmasses}
based on  the double seesaw mechanism 
and  additional flavor symmetry or/and  Grand Unification.

\section{Screening and renormalization group effect\label{sec:RG}}

In this section we will consider  effects of 
the radiative corrections~\cite{Grzadkowski:1987tf,Grzadkowski:1987wr,Casas:1999tp,Casas:1999ac} 
on the screening mechanism. The proportionality relation (\ref{eq:relation}) required 
for screening, holds presumably at the large scale of $M_S$ or above 
that. However, the couplings $Y_N$ and $Y_{\nu}$ have different 
gauge properties and their renormalization group running to the 
EW scale leads to different radiative 
corrections
which may destroy the  screening effect. 
On the other hand,  small  mismatch of the renormalized 
Yukawa matrices (partial screening) may explain some of the observed properties of the 
neutrino mass spectrum and mixings. 

In what follows we  will consider a minimal scenario - corrections due to the SM and MSSM interactions.  
Presence of other physics beyond the standard model can lead to additional renormalization effects. 

\subsection{Renormalization group effects}

\begin{figure}
\psfrag{MN1}{$M_1$}
\psfrag{MN2}{$M_2$}
\psfrag{MN3}{$M_3$}
\psfrag{Lambda}{$\Lambda$}
\psfrag{PHI}{$\braket{\phi}$}
\psfrag{RG3}{$(3)$}
\psfrag{RG2}{$(2)$}
\psfrag{RG1}{$(1)$}
\psfrag{RG0}{$(0)$}
\psfrag{mu}{$\mu$}
\includegraphics[width=\linewidth]{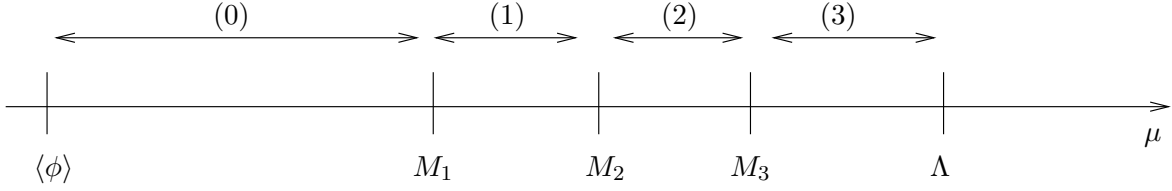}
\caption{The thresholds due to masses of the right-handed neutrinos 
         and the intervals of the RG running.}
\label{fig:Nthresholds}
\end{figure}

Here we will consider the renormalization group effects below a
certain scale $\Lambda$ which in turn is somehow below 
 the masses of singlets $M_{S}$: 
\begin{equation}
 M_{i} \ll  \Lambda \leq M_{S}, 
\end{equation}
here $M_{i}$ are the masses of RH neutrinos (see Fig.~\ref{fig:Nthresholds}).  
We consider $M_S$ as the mass matrix of singlets at $\Lambda$,  so that possible 
renormalization effects on $M_S$  due to the couplings  $Y_{N}$ are included.

Let us stress that for the hierarchical Yukawa matrix $Y_\nu$, and consequently $Y_N$, 
the mass spectrum of the RH neutrinos is  (in general) strongly hierarchical: 
$M_N \sim Y_\nu^2$. Therefore effects of the RG running  between different 
mass thresholds is crucial~\cite{King:2000hk,King:2000ce,Antusch:2002rr,Stech:2003sb,Antusch:2005gp}.    

Let us introduce the effective operator $\EFT{n}{O_M}$ which generates   
neutrino masses in the basis $(\nu, N)$ 
\begin{equation}
\mathscr{L} = - (\nu^T, N^T)\EFT{n}{O_M}(\nu^T, N^T)^T \; .
\end{equation}
The superscript $\left(n\right)$ designates the number of right-handed 
neutrinos which are not decoupled at a given energy scale, that is, neutrinos in 
the effective theory. This superscript will denote also a range of 
RG-running with a given number of RH neutrinos,  and we use also the 
notation $\EFT{n-m}{Z} \equiv \EFT{n}{Z}\EFT{n+1}{Z}\dots\EFT{m-1}{Z}\EFT{m}{Z}$.

Below  the scale $\Lambda$ the singlets $S$ are integrated out, 
and the effective operator $\EFT{3}{O_M}$ can be written as 
\begin{equation}
  \EFT{3}{O_M}(\Lambda) =  
  \left(\begin{array}{cc}
      0 & \EFT{3}{Y_\nu}\phi\\
      \EFT{3}{Y_\nu^T} \phi &  \EFT{3}{M_N} \\
    \end{array}\right)\; .
\end{equation}
Here 
\begin{equation}
\label{eq:RHmasses3}   
 \EFT{3}{M_N} = -\braket{\sigma}^2\EFT{3}{Y_{N}^T} M_{S}^{-1}\EFT{3}{Y_N}    
\end{equation}
is the mass matrix of the three RH neutrinos at the scale $\Lambda$,
and $\EFT{3}{Y_\nu}$ is the matrix of Yukawa couplings at $\Lambda$. 

\begin{figure}
  \centering
  \subfigure[Higgs
  self-coupling]{$\CenterObject{
\fmfframe(0,0)(0,0){    
\begin{fmfgraph*}(135,75)
\fmfleft{l1,l2}
\fmfright{r1,r2}
\fmf{fermion,label=$L$,label.side=right}{l1,v}
\fmf{fermion,label=$L$,label.side=left}{r1,v}
\fmf{scalar,label=$\phi$,label.side=right}{l2,w}
\fmf{scalar,label=$\phi$,label.side=left}{r2,w}
\fmf{scalar,left=1,tension=.3,label=$\phi$}{w,v}
\fmf{scalar,right=1,tension=.3,label=$\phi$}{w,v}
\fmfdot{w}
\fmfv{decor.shape=square,decor.filled=.5}{v}
\end{fmfgraph*}}}$}
\hspace{2cm}
  \subfigure[Gauge interactions]{$\CenterObject{
\fmfframe(0,0)(0,0){    
\begin{fmfgraph*}(135,75)
\fmfleft{l1,l2}
\fmfright{r1,r2}
\fmf{fermion,label=$L$,label.side=right}{l1,v}
\fmf{fermion,label=$L$,label.side=left}{r1,v}
\fmf{scalar,label=$\phi$,label.side=right}{l2,w1}
\fmf{scalar,label=$\phi$,label.side=left}{r2,w2}
\fmf{scalar,tension=.5,label=$\phi$,label.side=right}{w1,v}
\fmf{scalar,tension=.5,label=$\phi$,label.side=left}{w2,v}
\fmf{boson,tension=.5}{w1,w2}
\fmfdot{w1,w2}
\fmfv{decor.shape=square,decor.filled=.5}{v}
\end{fmfgraph*}}}$}
  \caption{The d=5 operator vertex corrections. Shown are additional 
           divergent diagrams in the effective theory.}
  \label{fig:divdiagrams}
\end{figure}
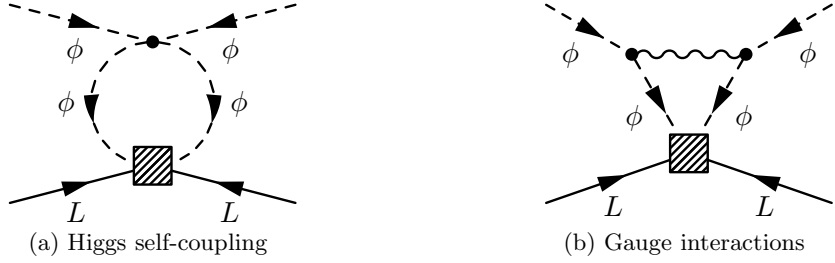

The effect of the RG evolution can be split in effects coming from the
renormalization of the wave functions and the vertex corrections.
It turns out, that the RG corrections can be factorized in the leading log (LL)  
approximation.  So,  in general, 
the renormalizations of $Y_\nu$, $M_N$ and $m_\nu \phi^2/\braket{\phi}^2$ 
(operator of the light neutrino masses) are  given by 
\begin{align}
  Y_\nu&\xrightarrow{\mathrm{RG}}Z_\mathrm{ext}^T Y_\nu Z_N\\
  M_N&\xrightarrow{\mathrm{RG}}Z_N^T M_N Z_N \\
  m_\nu \phi^2/\braket{\phi}^2& \xrightarrow{\mathrm{RG}}Z_\mathrm{ext}^T m_\nu
  \phi^2/\braket{\phi}^2 Z_\mathrm{ext} Z_\kappa\; . 
\label{eq:last}
\end{align}
Here $Z_\mathrm{ext}$ combines the renormalization effect of the left-handed doublets
$L$, the Higgs doublet $\phi$, and the vertex correction to $Y_\nu$; 
$Z_N$ denotes the wave function renormalization effect of the RH neutrinos $N$.
In order to simplify the presentation, we define the wave function 
renormalization so that the usual powers of $1/2$ 
factors are absent.  Eq.~(\ref{eq:last}) describes renormalization of the
effective dimension d=5 operator which appears after decoupling 
(integration out) of the corresponding RH neutrino. Apart from renormalization 
of the wave functions and vertices which exist in the SM model this 
operator has additional vertex corrections given by the diagrams in 
Fig.~\ref{fig:divdiagrams}. The RG effect due to these diagrams denoted 
by $\EFT{n}{Z_\kappa}$  plays a crucial role in the discussion of the 
stability of screening
These d=5 operator corrections are absent in the supersymmetric 
version of theory due to the non-renormalization 
theorem~\cite{Grisaru:1979wc,Seiberg:1993vc}. 
%

We describe the RG effects in the effective theory, where the heavy 
RH neutrinos are decoupled 
successively\footnote{The running between mass thresholds of RH neutrinos has
  been treated analytically in the approximation of strongly hierarchical and
  diagonal Yukawa matrix in ref.~\cite{Stech:2003sb}. Here we present a general consideration 
required for our approach.  
}~\cite{King:2000hk,King:2000ce,Antusch:2002rr,Stech:2003sb,Antusch:2005gp} 
as depicted in Fig.~\ref{fig:Nthresholds}. In each step (interval between 
mass thresholds) we first calculate the RG correction to the matrices.
We diagonalize the resulting matrices at the lower end of the interval, {\it i.e.}, 
at $\mu=M_i$ and then decouple $N_i$, $(i=3,2,1)$.
We will denote the renormalization factors in the leading log approximation 
by $\EFT{n}{Z}=1+\EFT{n}{\delta Z}$. 
This notation is also used for parameters of the effective theory. 
The renormalization factors in the extended (by RH neutrinos) SM and the MSSM are given in appendix~\ref{sec:renfactors}.

Let us describe the main steps of the renormalization procedure, which is 
especially simple in the basis where matrix $Y_\nu$ is diagonal. 
(This basis, however, does not coincide with the flavor basis.)

\noindent
1). The RG evolution between $\Lambda$ and $M_{3}$ yields the operator 
$\EFT{3}{O_M}$ at $M_3$
\begin{equation}
\label{eq:oper3}
\EFT{3}{O_M}(M_3) =  \left(\begin{array}{cc}
      0  & ~~~\EFT{3}{Z_\mathrm{ext}^T} \EFT{3}{Y_\nu}\phi
      \EFT{3}{Z_N} \\
      ... & ~~~\EFT{3}{Z_N^T} \EFT{3}{M_N} \EFT{3}{Z_N} \\
    \end{array}\right)\; .
\end{equation}
Performing a rotation of the RH neutrinos $N = \EFT{3}{U_N} N'$
we reduce the  renormalized RH neutrino mass matrix to the form
\begin{equation}
\label{eq:matrix3}
  \EFT{3}{U_N^T}\EFT{3}{Z_N^T} \EFT{3}{M_N} \EFT{3}{Z_N}\EFT{3}{U_N}=
  \left(
    \begin{array}{cc}
      \EFT{2}{M_N} & 0 \\
      0 & M_3
  \end{array}
\right)\; ,
\end{equation}
where $\EFT{2}{M_N}$ is $2\times2$ (in general off-diagonal) mass matrix. 
Let us split the $3\times 3$ Dirac type  Yukawa coupling matrix in (\ref{eq:oper3}) 
after this rotation into two parts as 
\begin{equation}
\EFT{3}{Z_\mathrm{ext}^T}  \EFT{3}{Y_\nu} \EFT{3}{Z_N}
  \EFT{3}{U_N} \equiv \left(\begin{array}{cc}\EFT{2}{Y_\nu}, & y_3
    \end{array}\right), 
  \end{equation}
where $y_3$ is the 3 component column of the Yukawa couplings of $\nu$ and $N_3$,  and 
$\EFT{2}{Y_\nu}$ is the rest $3\times2$ sub-matrix.   
Then in the rotated basis the operator (\ref{eq:oper3}) can be 
written  as 
\begin{equation}
\label{eq:oper3rot}
\EFT{3}{O_M}(M_3) =   \left(\begin{array}{cc}
      0 & \begin{array}{cc}\EFT{2}{Y_\nu}\phi  & y_3\phi
    \end{array} \\
      ... &  
    \begin{array}{cc}
      \EFT{2}{M_N} & 0 \\
      0 & M_3
  \end{array}\\
    \end{array}\right).
\end{equation}
Below the scale $M_3$ the neutrino $N_3$ is integrated out and from 
(\ref{eq:oper3rot}) we obtain
\begin{equation}
\EFT{2}{O_M}(M_3) = \left(\begin{array}{cc}
      -y_3 M_3^{-1}y_3^T\phi^2& \EFT{2}{Y_\nu}\phi \\
      ... &   \EFT{2}{M_N}\\
    \end{array}\right)\; .
\end{equation}
Notice that the $d = 5$ operator is formed in 1-1 block  due to 
decoupling of $N_3$. \\ 

\noindent
2). Let us consider the  RG running in the interval 
$M_2 - M_3$. Similarly to the first step we can write the operator  $O_M$  at 
the scale $M_2$ (threshold of $N_2$) as 
\begin{equation}
 \EFT{2}{O_M}(M_2) =  \left(\begin{array}{cc}
      -\EFT{2}{Z_\mathrm{ext}^T} \EFT{2}{Z_\kappa} y_3
  M_3^{-1}y_3^T \phi^2\EFT{2}{Z_\mathrm{ext}}
  & ~~~\EFT{2}{Z_\mathrm{ext}^T} \EFT{2}{Y_\nu}\phi\EFT{2}{Z_N}  \\
      ... &   ~~~\EFT{2}{Z_N^T}\EFT{2}{M}\EFT{2}{Z_N}\\
    \end{array}\right)\; .
\end{equation}
Here we have included the corrections $\EFT{2}{Z_\kappa}$ to the d=5 operator. 

By applying the rotation $N' = \EFT{2}{U_N} N''$ the renormalized mass matrix 
of the RH neutrinos is diagonalized: 
\begin{equation}
\label{eq:matrix2}
  \EFT{2}{U_N^T}\EFT{2}{Z_N^T}\EFT{2}{M}\EFT{2}{Z_N}\EFT{2}{U_N} \equiv
  \left(\begin{array}{cc}
 \EFT{1}{M_N} & 0 \\
  0 & M_2 \\
\end{array}\right). 
\end{equation}
The renormalized Yukawa matrix is then split as  
\begin{equation}
 \EFT{2}{Z_\mathrm{ext}^T} \EFT{2}{Y_\nu}\EFT{2}{Z_N}
      \EFT{2}{U_N} \equiv 
\left(\begin{array}{cc} \EFT{1}{Y_\nu}, & y_2 \end{array}\right),\; 
\end{equation}
where $\EFT{1}{Y_\nu}$ and $y_2$ are two component columns.
Decoupling the second neutrino $N_2$ we obtain  
\begin{equation}
\EFT{1}{O_M}(M_2) =  \left(\begin{array}{cc}
      -\EFT{2}{Z_\mathrm{ext}^T} \EFT{2}{Z_\kappa} 
y_3 M_3^{-1}y_3^T \phi^2 \EFT{2}{Z_\mathrm{ext}} - y_2 M_2^{-1}y_2^T \phi^2 & ~~~\EFT{1}{Y_\nu}\phi\\
      ... &   ~~~\EFT{1}{M_N}\\
    \end{array}\right).
\end{equation}\\

\noindent
3). Running the matrix down to the lowest see-saw scale 
$M_1$ and integration out $N_1$ yields
\begin{equation}
\begin{split}
\EFT{0}{O_M}(M_1)  = &-\EFT{1-2}{Z_\mathrm{ext}^T}
\EFT{1-2}{Z_\kappa} 
y_3 M_3^{-1}y_3^T\phi^2\EFT{1-2}{Z_\mathrm{ext}}\\
  &-\EFT{1}{Z_\mathrm{ext}^T} \left[\EFT{1}{Z_\kappa}y_2
  M_2^{-1}y_2^T + 
\EFT{1}{Y_\nu} \EFT{1}{M_N^{-1}}\EFT{1}{Y_\nu^T}\right]
\phi^2\EFT{1}{Z_\mathrm{ext}}\; .
\end{split}
\end{equation}\\

\noindent
4). Finally, evolving $\EFT{0}{O_M}(M_1)$    
down to the EW scale, we obtain (after $\phi$ develops a VEV) the mass 
matrix of light neutrinos
\begin{equation}
\label{eq:main1}
m_\nu=-\braket{\phi}^2\EFT{0-3}{Z_\mathrm{ext}^T} \EFT{3}{Y_\nu} 
\EFT{3}{Z_N} \EFT{3}{U_N}
\left(\begin{array}{cc}
K_{12} & 0 \\
  0 &  \frac{\EFT{0-2}{Z_{\kappa}}}{M_3}  \\
\end{array}\right)
\EFT{3}{U_N^T} \EFT{3}{Z_N^T} \EFT{3}{Y_\nu^T}\EFT{0-3}{Z_\mathrm{ext}}\; ,
\end{equation}
with  
\begin{equation}
K_{12} \equiv  \EFT{2}{Z_N} \EFT{2}{U_N}
      \left(\begin{array}{cc}
  \frac{\EFT{0}{Z_\kappa}}{\EFT{1}{M_N}} & 0\\
          0 & \frac{\EFT{0-1}{Z_\kappa}}{M_2}\\
        \end{array}\right) \EFT{2}{U_N}^T \EFT{2}{Z_N^T} .
\end{equation} \\
This expression can be presented in a simpler and transparent way. 
Using definitions of matrices  $\EFT{2}{U_N}$  and  
$\EFT{3}{U_N}$ in eqs. (\ref{eq:matrix2}, \ref{eq:matrix3}) 
we can rewrite $m_\nu$  as 
\begin{equation}
\label{eq:lightnu}
m_\nu=-\braket{\phi}^2 Z_\mathrm{ext}^T  
\left[\EFT{3}{Y_\nu} 
    X_N \EFT{3}{M_N^{-1}} \EFT{3}{Y_\nu^T}\right]Z_\mathrm{ext} \; ,
\end{equation}
where 
\begin{equation}
\label{eq:xN}
  X_N\equiv  \EFT{3}{Z_N}\EFT{3}{U_N} \EFT{2}{Z_N'} \EFT{2}{U_N'}
 Z_\kappa \EFT{2}{U_N^{'\dagger}}\EFT{2}{Z_N^{'-1}}
\EFT{3}{U_N^\dagger} \EFT{3}{Z_N^{-1}}
\end{equation} 
describes the RG effects due to the running between the thresholds. 
Here 
\begin{equation}
\label{eq:3dmatrix}
\EFT{2}{Z_N'} \equiv  
  \left(\begin{array}{cc}\EFT{2}{Z_N} & 0 \\ 0 & 1 \\\end{array}\right),~~~
\EFT{2}{U_N'} \equiv 
  \left(\begin{array}{cc} \EFT{2}{U_N} & 0 \\ 0 & 1 \\\end{array}\right),~~~
 \end{equation} 
and 
\begin{equation}
Z_\kappa \equiv 
\diag\left(\EFT{0}{Z_\kappa},  \EFT{0-1}{Z_\kappa},  \EFT{0-2}{Z_\kappa}\right) 
\end{equation}
is the matrix of effective d=5 operator corrections (fig.~\ref{fig:divdiagrams}).  
The expressions for $Z_\mathrm{ext}$ are given in appendices~\ref{sec:appone}
and \ref{sec:apptwo}.
Formulas (\ref{eq:main1}) and (\ref{eq:lightnu}) for $m_\nu$ are our main
results which we will analyze in the following sections.

\subsection{Stability of screening}

Inserting the expression for the matrix $M_N$, eq.~(\ref{eq:RHmasses3}), 
into eq.~(\ref{eq:lightnu}) we obtain 
\begin{equation}
\label{eq:lightnu2}
    m_\nu=  \frac{\braket{\phi}^2}{\braket{\sigma}^2 }  
Z_\mathrm{ext}^T\left[\EFT{3}{Y_\nu} X_N \EFT{3}{Y_N^{-1}} M_S \EFT{3}{Y_N^{T -1}}
\EFT{3}{Y_\nu^T}\right]Z_\mathrm{ext} \; .
\end{equation}
If the equality (\ref{eq:relation}) is satisfied, the neutrino mass 
matrix becomes
\begin{equation}
\label{eq:numass1}    
m_\nu =  \frac{\braket{\phi}^2}{\braket{\sigma}^2 }  
{Z_\mathrm{ext}^T \left[\EFT{3}{Y_\nu} X_N \EFT{3}{Y_\nu^{-1}} M_S \right] Z_\mathrm{ext}}\; .
\end{equation}
Therefore screening is reproduced and the dependence of $m_\nu$ on 
the Yukawa (Dirac) couplings disappears if $X_N = I$.
The expression (\ref{eq:numass1}) coincides with that in
(\ref{eq:screening}) up to external renormalization.  
In turn, according to eq.~(\ref{eq:xN}) the equality $X_N = I$ holds provided that 
$Z_{\kappa}  = I$,  that is, when the d=5 operator corrections 
are absent. This is automatically satisfied in the supersymmetric 
theory, but the corrections are present in the SM and its 
non-supersymmetric extensions. 

Note that the d=5 operator corrections are due to the gauge interactions 
and self interactions of the Higgs boson, which are by themselves flavor 
blind. However, they influence the flavor structure of the light 
neutrino mass matrix due to difference of masses of the RH 
neutrinos and therefore different threshold effects.

\subsection{Beyond leading log approximation}

\begin{figure}\centering
$\CenterObject{
\fmfframe(0,0)(0,0){
\begin{fmfgraph*}(200,100)
\fmfleft{l}
\fmfright{r}
\fmf{fermion,label=$L$,label.side=left}{l,v1}
\fmf{fermion,label=$e_R;N$,label.side=right}{v1,v2}
\fmf{fermion,label=$L$,label.side=left}{v2,u}
\fmf{fermion,label=$L$}{r,w1}
\fmf{fermion,label=$e_R;N$,label.side=left}{w1,w2}
\fmf{fermion,label=$L$}{w2,u}
\fmfdot{v1,v2,w1,w2}
\fmfv{decor.shape=square,decor.filled=.5}{u}
\fmffreeze
\fmfright{s1,s2}
\fmfleft{k1,k2}
\fmf{scalar,left=.8,tension=.3,label=$\phi$}{v1,w2}
\fmf{scalar,right=.8,tension=.3,label=$\phi$}{w1,v2}
\fmf{scalar,label=$\phi$,label.side=left}{s1,u}
\fmf{scalar,label=$\phi$,label.side=right}{k1,u}
\end{fmfgraph*}}}$
  \caption{Examples of the two loop diagram which destroy 
      factorization of the vertex corrections to the effective neutrino mass matrix.}
    \label{fig:MnuTwoL}
  \end{figure}
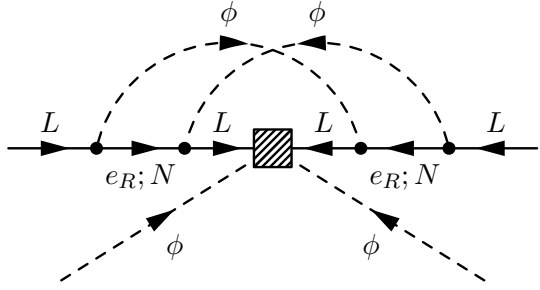

Beyond the leading log (LL) approximation the RG contributions still factorize 
except for the d=5 operator corrections. Thus,  our analysis does hold
beyond  LL in the MSSM, but it does not apply in the SM due 
to diagrams like in Fig.~\ref{fig:MnuTwoL}.
However, the 2 loop contributions to the external renormalization are smaller 
than the 1~loop corrections by a factor of $y^2/16\pi^2 \le 0.01$ and 
the 1~loop threshold corrections are not enhanced by large logarithms.

For certain structures of $M_S$ (see sec.~\ref{sec:applications})
the additional 2~loop diagrams lead to corrections to the renormalization 
of the effective neutrino mass operator which could be comparable to 
the 1~loop corrections. However, assuming the same hierarchy in the neutrino 
Yukawa couplings as in the up-type quark Yukawa couplings, these 
contribution are further suppressed since the heaviest right-handed 
neutrino is already integrated out. Therefore all 2~loop contributions 
in the effective theory are suppressed by $y^2_2/16\pi^2 \le 10^{-6}$ 
compared to the 1~loop contributions. Altogether higher loop contributions 
are less than $10\%$ of the 1~loop corrections and will be neglected in 
the following.

\subsection{Effects of vertex corrections}

In order to study the d=5 operator corrections  in the 
(non-supersymmetric) standard model in more details we introduce the 
matrix $V_N$ which diagonalizes the RH neutrino mass matrix at $\Lambda$: 
\begin{equation}
\label{eq:invert}
V_N^T \EFT{3}{M_N} {V_N} = 
\braket{\sigma}^2 {V_N^T} \EFT{3}{Y_N^T} {M_S^{-1}} \EFT{3}{Y_N} {V_N} 
= {D_N} \equiv \diag(M_1, M_2, M_3).
\end{equation}
In the lowest order approximation:   $\EFT{2}{Z_N} = \EFT{3}{Z_N} = 1$,   
and according to (\ref{eq:matrix3}) and 
(\ref{eq:matrix2}) we obtain    
\begin{equation}
V_N =  \EFT{3}{U_N} \EFT{2}{U_N'}. 
\end{equation}
Therefore the matrix $X_N$ (\ref{eq:xN}) can be rewritten in the form   
\begin{equation}
  X_N = V_N Z_{\kappa} V_N^{\dagger} = 
   I + {V_N} \delta Z_{\kappa} V_N^{\dagger},  
\end{equation}
where 
$$
\delta Z_{\kappa} \equiv Z_{\kappa} - I. 
$$
Plugging this expression for $X_N$ in (\ref{eq:lightnu2}) we find 
\begin{equation}
\label{eq:tilde1}
{m}_\nu \equiv 
\frac{\braket{\phi}^2}{\braket{\sigma}^2 }
Z_\mathrm{ext}^T  \left[I + \EFT{3}{Y_\nu} V_N \delta Z_{\kappa} V_N^{\dagger} 
\EFT{3}{Y_\nu}^{-1} \right]M_S Z_\mathrm{ext} . 
\end{equation}
Finally, using \eqref{eq:invert} we can rewrite ${m}_\nu$ as  
\begin{equation}
\label{eq:tilde2}
{m}_\nu \equiv 
\braket{\phi^2} Z_\mathrm{ext}^T    \left[\frac{1}{\braket{\sigma}^2} M_S + 
\EFT{3}{Y_\nu} V_N (\delta Z_{\kappa} D_N^{-1}) V_N^T 
\EFT{3}{Y_\nu^T}\right] Z_\mathrm{ext}. 
\end{equation}
According to this expression the effects of d=5 operator corrections 
are reduced to renormalization of masses of the RH neutrinos since  
$\delta Z_{\kappa}$ is diagonal.

\section{Applications of the Dirac structure screening
\label{sec:applications}}

Here we apply the results obtained in the previous section to several 
phenomenologically interesting structures  of $M_S$. 
We study effects of the radiative corrections on the light neutrino mass matrix. 
The matrix $M_S$ will be defined in the basis where the equality of the Yukawa 
couplings~(\ref{eq:relation}) is fulfilled. 

In the previous sections we have found $m_\nu$ in the basis where the 
neutrino Yukawa coupling matrix $Y_\nu$ is diagonal. Now we will 
discuss $m_\nu^f$ -  the neutrino mass matrix in the flavor basis
where the charge lepton mass matrix $Y_e$ is diagonal. It is related 
to $m_\nu$ as  
\begin{equation}
\label{eq:flavb}
m_\nu^f = U_e^T m_\nu U_e, 
\end{equation}
where $U_e$ is the transformation of left handed charged lepton 
components which diagonalizes the matrix $Y_e$ at the electroweak
scale. The radiative corrections to $Y_e$ are in general small due 
to the strong mass hierarchy. 

\subsection{Quasi-degenerate neutrino spectrum\label{sec:quasidegen}}

Let us first consider  $M_S$ which is proportional to the unit 
matrix at $\Lambda$, {\it i.e.}, 
\begin{equation}
\label{eq:unit}
M_S = M_S^0 I\; .
\end{equation}
This choice is apparently basis independent and we can take 
therefore $Y_{\nu} = Y_{N} = \diag(y_1, y_2, y_3)$. The 
right-handed neutrino mass matrix is diagonal and strongly hierarchical: 
\begin{equation}
  M_N=-Y_N^T M_S^{-1} Y_N\braket{\sigma}^2
     =\frac{\braket{\sigma}^2}{M_S}\diag\left(y_1^2,\,y_2^2,\,y_3^2\right)\; .
\end{equation}
Therefore $V_N = I$ and we find 
\begin{equation}
\label{eq:tilde1a} 
m_\nu^f \equiv 
\frac{\braket{\phi}^2}{\braket{\sigma}^2 } M_S^0
 U_e^T  Z_\mathrm{ext}^T \left[I + \delta Z_{\kappa} \right] Z_\mathrm{ext} U_e. 
\end{equation}
The corrections are also diagonal\footnote{We assume a strong hierarchy in
  $Y_\nu$ and use $Y_\nu\sim Y_u$ for the numerical estimates.} 
\begin{equation}
\delta Z_{\kappa}  =  \left[\exp\left(\mathcal{A}~  \diag\left(0,\,\ln\frac{y_2^2}{y_3^2},\,
\ln\frac{y_1^2}{y_3^2}\right)\right)-I\right]\sim \Ord{0.1}\; ,\label{eq:effvcor}
\end{equation}
where 
\begin{equation}
\label{eq:loopf}
\mathcal{A} \equiv \frac{1}{16\pi^2}\left(\lambda+\frac{9}{10}g_1^2+ 
\frac{3}{2}g_2^2\right). 
\end{equation}
This leads to splittings of the light neutrino masses which would 
be degenerate otherwise. 

Note that the external corrections (due to the wave function renormalization 
of the left-handed leptons, eq.~\eqref{eq:ZL}, and the vertex corrections 
to the neutrino Yukawa couplings, eq.~\eqref{eq:ZYnu}), are described 
in general by off-diagonal matrices due to the mismatch of $Y_e$ 
and $Y_\nu$ structures. 
As the charged lepton Yukawa couplings are also strongly hierarchical, 
the largest flavor dependent correction is the one  to the 3-3 element. Neglecting the
off-diagonal entries, it can be estimated as 
\begin{equation}
-2\frac{y_\tau^2}{16\pi^2}\ln\frac{\braket{\phi}}{\Lambda} - 4\frac{y_3^2}{16\pi^2}\ln
\frac{M_3}{\Lambda}\sim\Ord{0.1} ,  
\end{equation}
where the second term (due to the neutrino Yukawa coupling) dominates.
It has the same order of magnitude as the correction due to the d=5 operator 
renormalization in eq.~(\ref{eq:effvcor}). 

Let us now comment on a possibility to explain the neutrino data. 
In the non-supersymmetric version the mass split, $\Delta m$, 
generated by the d=5 operator corrections, $\Delta m = m_0 \delta Z$, leads 
to $\Delta m^2 = 2 m_0 \Delta m = 2 m_0^2\delta Z = (2 - 8)\cdot
10^{-3}$~eV$^2$ for the overall scale $m_0 = (0.1 - 0.2)$~eV. This can reproduce 
the atmospheric mass split, but it is too large for the solar mass split.  
The ratio of solar and atmospheric mass squared differences, 
\begin{equation}
  \frac{\Delta m_\mathrm{21}^2}{\Delta
  m_\mathrm{32}^2}~~\approx~~\frac{m_2-m_1}{m_3-m_2}\sim\Ord{1}\; ,
\end{equation}
does not fit the observations.  The external corrections do not improve the  situation either. 
Therefore some other (non-radiative)
contribution is required to compensate the 1-2 mass split. Mixings can
also be generated by small (non-radiative) corrections. 

In the supersymmetric version we have $\delta Z_\kappa = 0$ and $Z_{Y_{\nu}} = I$, so that the mass splitting is produced by the external 
renormalization only:  
\begin{equation}
\label{eq:massde}
m_\nu^f =  \frac{\braket{\phi}^2}{\braket{\sigma}^2 } M_S^0 U_e^T Z_\mathrm{ext}^T U_e
Z_\mathrm{ext} \; . 
\end{equation}
In the flavor basis we obtain the mass split due to Yukawa couplings coming from
the external renormalization: 
\begin{equation}
\label{eq:corr2}
 \exp\left[- \frac{1}{8\pi^2} \diag(y_e^2,~y_\mu^2,~ y_\tau^2)
\ln\frac{\braket{\phi}}{\Lambda}\right] \; ,
\end{equation}
where the neutrino Yukawa couplings are neglected.
This can provide the atmospheric mass split and the mixings
should be generated again by correction to the zero order structure.

Next,  we consider for $M_S$ the ``triangle'' structure 
\begin{equation}
\label{eq:triang}
M_S =   M_S^0\left(\begin{array}{ccc}1
    &0&0\\0&0&1\\0&1&0\\\end{array}\right)
\end{equation}
in the basis where the neutrino Yukawa matrices are diagonal. 
In lowest order it produces a degenerate mass spectrum and maximal 
2-3 mixing of the light neutrinos. This matrix leads to a spectrum 
of RH neutrinos with  two heavy degenerate states and one
relatively light state: 
\begin{equation}
M_{1N} = \frac{\braket{\sigma}^2 y_1^2}{M_S^0}, \quad\quad
M_{2N} = - M_{3N} =  \frac{\braket{\sigma}^2 y_2 y_3}{M_S^0}.  
\end{equation}
The renormalization interval $``(2)"$ (see fig.~\ref{fig:Nthresholds}) is 
therefore absent and the matrix of d=5 operator corrections 
can be written as 
\begin{equation}
\label{effvcor1}
\delta Z_{\kappa} = \left[\exp\left(\delta \EFT{1}{Z_\kappa}\right) -1\right] 
\diag(0, 1, 1)\; , 
\quad\quad \delta \EFT{1}{Z_\kappa} =  \mathcal{A} \ln\frac{y_1^2}{y_2 y_3} \; , 
\end{equation}
where $\mathcal{A}$ is defined in eq.~(\ref{eq:loopf}). The state $N_1$ 
decouples and maximal mixing is realized in the 2-3 block of $V_N$.
Using this feature and eq.~(\ref{effvcor1}) we find from 
eq.~(\ref{eq:tilde2}) 
\begin{equation}
\label{eq:tilde3}
m_\nu^f \equiv 
Z_\mathrm{ext}^T \frac{\braket{\phi}^2}{\braket{\sigma}^2 } M_S^0 U_e^T \EFT{0}{Z_\kappa}
\left(\begin{array}{ccc}
    1&0&0\\
0 & 0 & 1 - \delta \EFT{1}{Z_\kappa} \\
0 & 1 - \delta \EFT{1}{Z_\kappa} & 0\\
\end{array}\right)Z_\mathrm{ext} U_e\; .
\end{equation}
Therefore the d=5 operator corrections do  not destroy the triangular 
structure, but they lead to mass splits between the degenerate 
pair and the isolated state: 
\begin{equation}
\label{eq:massspl}
\frac{\Delta m}{m} = \delta \EFT{1}{Z_\kappa} .  
\end{equation}
In the supersymmetric version $\delta \EFT{1}{Z_\kappa} = 0$,  so that the 
original ``triangle'' structure is renormalized by the external
corrections only. In this case, one also needs  perturbations 
of the original screening structure in order to get the correct mixings 
and mass split. 

As a third possibility we consider for $M_S$ the  ``triangle'' structure 
which leads to a degenerate spectrum and maximal 1-2 mixing: 
\begin{equation}
\label{eq:mat-t2}
M_S = M_S^0
  \left(\begin{array}{ccc}
      0 & 1 & 0 \\
      1 & 0 & 0 \\
      0 & 0 & 1\\
    \end{array}\right)\; . 
\end{equation}
Similar considerations as above results in  the mass spectrum of RH neutrinos 
with two light degenerate states and an isolated heavier state:
\begin{equation}
M_{1N}  = - M_{2N} = \frac{\braket{\sigma}^2 y_1 y_2}{M_S^0}, ~~~~~~
M_{3N} =  \frac{\braket{\sigma}^2 y_3^2}{M_S^0} \;.
\end{equation}
For the light neutrinos we find 
\begin{equation}
\label{eq:tilde3a}
m_\nu \equiv 
Z_\mathrm{ext}^T \frac{\braket{\phi}^2}{\braket{\sigma}^2 } M_S^0 U_e^T \EFT{0}{Z_\kappa}
\left(\begin{array}{ccc}
0 & 1 & 0\\
1 & 0 & 0 \\
0 & 0 &1 -  \delta \EFT{2}{Z_\kappa} \\
\end{array}\right)Z_\mathrm{ext} U_e \; , 
\end{equation}
where 
\begin{equation}
\quad\quad\delta \EFT{2}{Z_\kappa} =  \exp\left(\mathcal{A} \ln\frac{y_1y_2}{y_3^2}\right) - 1\; .
\end{equation}
The corrections due to running of the d=5 operator are of the same order as
in the previous case. The mass split 
\begin{equation}
\Delta m_{32}^2=2 m_0 \Delta m_{32}=-2m_0^2 
\delta\EFT{2}{Z_\kappa}= (2-8)\cdot 10^{-3}\eV^2\; ,
\end{equation}
for $m_0= (0.08 - 0.16)~ \eV$  can  explain the atmospheric 
neutrino data. The external renormalization contributes in the same way as for 
$M_S\propto I$.

In the supersymmetric version the ``triangle'' structure is preserved. 
Therefore the original matrix $M_S$ as given in  eq.~(\ref{eq:mat-t2}) 
should be perturbed in order to produce phenomenological acceptable mixings.

Let us comment on the possibility to generate the lepton asymmetry of the Universe via 
the CP-violating decay of the lightest right handed neutrino(s)~\cite{Fukugita:1986hr}. 
In the case of diagonal $M_S$ the mass of the lightest neutrino $N_1$ is in the TeV range. 
So, according to~\CiteBoundOnMN\ the produced asymmetry is too small to explain
the Baryon asymmetry via sphaleron effect. In contrast,  in the case of a triangular 
matrix $M_S$ (\ref{eq:tilde3a}) there are two quasi-degenerate RH neutrinos
with masses $M\simeq 10^{6}~\GeV$, and  resonant leptogenesis can produce the 
required asymmetry. 

\subsection{Perturbations of $M_S$\label{sec:perturbMS}}

Let us  consider perturbations of the structure of $M_S$ 
(which can be required  by phenomenology)
and effect  radiative corrections on these perturbed structures. 

As an example we take the matrix 
\begin{equation}
\label{eq:text2}
M_S =   M_S^0\left(\begin{array}{ccc}
1 & 0 & 0\\
0 & x & 1\\
0 & 1 & 0\\
\end{array}
\right) \;   
\end{equation}
with $x$ being a free  parameter. 
Now the second and third neutrinos are no longer degenerate
and the renormalization factor $\EFT{2}{Z_\kappa}$ in the interval $``(2)"$ 
between their masses appears. 
Approximating $\EFT{n}{Z_\kappa}$  by $1+\mathcal{A}\ln(M_n/M_{n+1})$ 
we obtain for the light neutrinos
\begin{equation}
\label{eq:tilde5}
m_\nu^f =  \frac{\braket{\phi}^2}{\braket{\sigma}^2}M_S^0 U_e^T \EFT{0}{Z_\kappa}Z_\mathrm{ext}^T
\begin{pmatrix}
    1 & 0 & 0 \\
    ... & x\left(1+\mathcal{A}\right) m^\mathrm{th}_{22} &
    1+\mathcal{A} m^\mathrm{th}_{23}\\
    ... & ... & \mathcal{A} m^\mathrm{th}_{33}
  \end{pmatrix}
  Z_\mathrm{ext} U_e \; , 
\end{equation}
where the threshold dependent corrections, $m_{ij}^\mathrm{th}$, equal  
\begin{align} 
m^\mathrm{th}_{22}=&-3\ln\lambda+\left(\frac{1}{2}-\frac{\lambda^2}{x^2y}\right)\ln\frac{y-1}{y+1}~,\nonumber\\
m^\mathrm{th}_{23}=&-3\ln\lambda+\frac{1}{2y}\ln\frac{y-1}{y+1}~,\label{eq:PerturbMS}\\
m^\mathrm{th}_{33}=&\frac{1}{xy}\ln\frac{y-1}{y+1}~.\nonumber 
\end{align}
Here  $y\equiv\sqrt{1+4\left(\frac{\lambda}{x}\right)^2}$ and  
$\lambda \equiv y_2/y_3$. (The logarithms depend on the ratios on the RH neutrino masses $M_2/M_3$.)

The  nonzero 3-3 element is generated in eq.~(\ref{eq:tilde5}) 
by the radiative corrections.  
Furthermore, this element can be enhanced by small parameter $x$ in the denominator, provided that 
$\lambda$ is also small enough. 
Indeed, from eq.~\eqref{eq:PerturbMS} we find  explicitly
\begin{equation}
(m_\nu)_{33} =
\left\{\begin{aligned}
\frac{2\mathcal{A}}{x} \ln\frac{\lambda}{x}, \quad\quad x \gg \lambda \\
-\frac{1.26\mathcal{A}}{x},  \quad\quad x = 2 \lambda\\
-\frac{\mathcal{A}x}{2\lambda^2},  \quad\quad x \ll \lambda\\
\end{aligned}\right.
\end{equation}
Since  $\mathcal{A} \sim 10^{-2}$,  the 3-3 element can be of the order 1 or even more 
if, {\it e.g.},   $\lambda \ll x < 10^{-2}$.    
Thus, a quasi-degenerate $M_S$ with nearly 
maximal 2-3 mixing leads after (non-supersymmetric) renormalization 
group corrections to the hierarchical mass matrix $m_{\nu}$ with small mixing. 
The texture (\ref{eq:text2}) is not stable against quantum corrections, 
since the structure of $m_{\nu}$ strongly differs from the original 
structure of $M_S$. 

This example shows that radiative corrections 
can  substantially modify the original texture of $M_S$ in the 
light neutrino mass matrix for particular  $M_S$. 
In other words, radiative corrections 
may destroy screening.

Apparently the corrections are small if $\lambda \ll x \sim 1$. This corresponds to 
phenomenologically important case of the dominant 2-3 block: 
\begin{equation}
\label{eq:text4}
M_S =   M_S^0\left(\begin{array}{ccc}
\epsilon & 0 & 0\\
0 & x & 1\\
0 & 1 & x\\
\end{array}
\right) \;
\end{equation}
with $x \sim 1$ and $\epsilon \ll 1$.

In the supersymmetric version of model screening is stable, since there are no 
d=5 operator corrections due to non-renormalization theorem. 
Note also that exact off-diagonal structures of $M_S$ are stable, 
but small perturbations are unstable with respect to radiative corrections. 

\subsection{Neutrino symmetry}

The mass matrix $M_S$ of the singlets $S$ has no analogy in the quark 
sector, and general,  it is not related to the quark mass matrices. 
Furthermore,  $M_S$ is generated at a higher scale, than the GUT scale. 
Therefore it is possible that $M_S$ has certain symmetry which does not 
show up (or is broken) at lower scales. If screening is realized as 
discussed in this paper, this symmetry propagates immediately 
to the light neutrino sector. $M_S$ can therefore be the origin 
of a specific ``neutrino symmetry''  which is not seen in the quark 
and in the charged lepton sectors. So,   screening allows us to reconcile special 
``neutrino'' symmetry~\CiteMuTauSymm\ and the quark-lepton symmetry. 
 
The examples considered in sec.~\ref{sec:quasidegen} illustrate this possibility, since the 
matrices  (\ref{eq:unit},\ref{eq:triang},\ref{eq:mat-t2}) 
have certain symmetries. These symmetries are protected from large 
corrections in the supersymmetric case, while they can be strongly broken 
in the non-supersymmetric versions, or when some other interactions and fields beyond
the SM or MSSM exist. 

\subsection{Quark-lepton complementarity}

According to the quark--lepton complementarity 
relation~\CiteQLC, 
the lepton mixing is given by maximal mixing minus the quark mixing. 
This may imply an existence of  certain structure in the lepton sector 
which generates maximal (bi-maximal) mixing and the quark-lepton 
symmetry in some form. In models with  screening mechanism, 
the mass matrix $M_S$  
which has no analogy in the quark sector, 
can  be the origin of the bi-maximal mixing. Then  the CKM type mixing follows from
the charged lepton mass matrix which is related to the mass matrix of
the down quarks,  so that $U_e = U_{CKM}$. In the lowest order 
(without radiative corrections) we find from eq.~(\ref{eq:flavb}) 
\begin{equation}
\label{eq:compl}
m_\nu^f = 
\left[\frac{\braket{\phi}}{\braket{\sigma}}\right]^2 U_e^T M_S U_e =
 \left[\frac{\braket{\phi}}{\braket{\sigma}}\right]^2 U_{CKM}^T
 U_{bm}^* M_S^{diag}U_{bm}^\dagger U_{CKM}, 
\end{equation}
where $U_{bm} \equiv R_{23}^{m} \times R_{12}^{m}$ 
is the bi-maximal mixing matrix and $R_{ij}^{m}$ are 
the $\pi/4$-rotations  in the 
$({ij})$ planes.  So, the leptonic 
mixing matrix 
equals $U_{PMNS} = U_{CKM}^\dagger U_{bm}$. This  
realizes the so called 
``neutrino scenario'' which leads to deviations from the exact quark-lepton 
complementarity relation~\cite{Minakata:2004xt}. 

The bi-maximal mixing is produced by the mass 
matrix~\CiteBiMax
\begin{equation*}
  M_S^\mathrm{bimax}=\left(\begin{array}{ccc}
      D-C & B & -B \\
      ... & D & C\\
      ... & ...  & D\\
    \end{array}\right) \; ,
  \end{equation*}
where $B, C, D$ are arbitrary parameters. As we have found in section~\ref{sec:perturbMS} in 
the non-supersymmetric case, such a matrix may be  unstable with respect to 
radiative corrections: the d=5 operator corrections  
can, in particular, suppress the maximal 2-3 mixing.  
This can be easily seen for vanishing $B$ and small $D$, when  
$M_S$ reduces to  a triangular structure which is not stable under radiative 
corrections, as we have shown in sec.~\ref{sec:perturbMS}.

\subsection{Singular $M_S$\label{sec:singular}}

Let us consider the special case of  singular 
$M_S$,  $\det M_S = 0$,  which can be a consequence of 
certain symmetry in $S$ sector.  
Now  one can not use immediately
(\ref{eq:DoubleSeeSaw}) and  the whole double seesaw mass matrix should 
be considered. In what follows we will show that the tree-level 
mass matrix of light neutrinos is still proportional to $M_S$, that is, 
eq.~(\ref{eq:DoubleSeeSaw}) will hold even if $M_S$ is singular.
For this we will compare the light neutrino mass spectra in the lowest 
approximation found from the whole double seesaw matrix $\mathcal{M}$ 
and from the matrix $m_\nu$ after decoupling of the heavy degrees of 
freedom in eq.~(\ref{eq:DoubleSeeSaw}). 

According to 
eq.~(\ref{eq:DoubleSeeSaw}) the condition  $\det M_S = 0$ implies (at least one) 
zero eigenvalue in the spectrum of usual LH neutrinos. The same follows from the complete matrix. 
Indeed, 
$$
\det\mathcal{M} = - \left(\det Y\right)^2\det M_S = 0, 
$$ 
and hence, zero eigenvalue of $M_S$ leads to a massless eigenstate of  
$\mathcal{M}$. The non-zero eigenvalues of the matrix $m_\nu$ , $\xi_i$,  
coincide with the eigenvalues of the full matrix $\mathcal{M}$ up to 
corrections of the order $\braket{\phi}/\braket{\sigma}$. This can be 
seen by inserting $\xi_i$ in the characteristic polynomial of the 
complete matrix $\chi_{\mathcal{M}}\left[\lambda\right]$. 
The result is of order  
$\cal{O}$ $\left(\left(\braket{\phi}/\braket{\sigma}\right)^8 \right)\sim 0$. 
Hence, $\xi_i$ are to a very good approximation the eigenvalues of 
$\mathcal{M}$.

There are no other light states, because the expansion of the polynomial 
$$
\chi_{\mathcal{M}}\left[\lambda\right] \prod_i 
\left(\lambda - \xi_i\right)^{-1},
$$ 
in eigenvalues of the order $\braket{\phi}$ does not yield any new 
solutions.  All other eigenvalues are at least of the order  
$\Ord{\braket{\sigma}^2/M_S}$.

A peculiarity of the spectrum of $\mathcal{M}$ is the appearance of one 
heavy Dirac particle, if the eigenstate of $M_S$ with zero mass, $S$,  
couples to only one  right-handed neutrino $N$. This Dirac particle is 
formed by  $S$ and $N$. 

The mass spectrum can be easily obtained if 
$M_S=\diag\left(M_{S1},~ M_{S2},~ 0\right)$  in the basis where 
$Y_N=\diag\left(y_1,\,y_2,\,y_3\right)$. Apart from one zero mass
which corresponds mainly to   $\nu_3$, 
and two super heavy eigenvalues $M_{S1}$ and $M_{S2}$ for two singlets
$S$,   we find  
\begin{equation*}
m_1 =  M_{S1}\frac{\braket{\phi}^2}{\braket{\sigma}^2},~~
m_2 =  M_{S2} \frac{\braket{\phi}^2}{\braket{\sigma}^2}, ~~ 
M_1 = -\frac{y_1^2 \braket{\sigma}^2}{M_{S1}}, ~~
M_2 = -\frac{y_1^2 \braket{\sigma}^2}{M_{S2}}, ~~
M_{DS} = y_3 \braket{\sigma},   
\end{equation*}
that is,  two light neutrinos predominantly 
given by $\nu_{1,2}$ with masses $m_1$ and $m_2$, 
two heavy neutrinos mostly consisting of $N_{1,2}$ with masses $M_1$ and $M_2$
and one heavy  Dirac particle of the GUT scale mass $M_{DS}$ which is formed 
by $N_3$ and $S_3$. 
The light eigenstates are mainly composed of the left-handed neutrinos 
and the mixing with other neutral leptons  is the order 
$\Ord{\braket{\phi}/\braket{\sigma}}$. 

The coincidence of the spectrum of $m_\nu$ and the spectrum of light 
states of $\mathcal{M}$ is related essentially to the fact that the 
relation between $m_\nu$ and $M_S$ is linear, and the characteristic 
polynomial is linear in the eigenvalues for the non-degenerate case.
The same conclusion holds  for $M_S$ with two zero eigenvalues.

Let us consider the effect of radiative corrections for this singular case.
As long as all contributions to a Majorana mass matrix $m_\nu$ receive the same quantum corrections, the RG 
evolution does not generate non-zero masses from vanishing 
masses~\cite{Chankowski:2001mx}. 
However, between the mass thresholds of the RH neutrinos, 
there are two contributions from the decoupling of the RH neutrinos 
which are renormalized differently. One contribution is due to the d=5 operator 
decoupled and the other is due to the contribution of the RH neutrinos which are not
decoupled yet ($Y_\nu M_N^{-1} Y_\nu^T$) in the intervals $M_2 - M_3$ and $M_1 - M_2$. Hence, 
the generated mass is proportional to the additional renormalization factor $\delta Z_\kappa$ 
from the d=5 operator between the thresholds and the mismatch between 
the two mass contributions, {\it i.e.} the deviation of the unitary matrix 
transforming from the eigenbasis of the d=5 operator to the eigenbasis 
of $Y_\nu M^{-1} Y_\nu^T$ between the thresholds from 
the unit matrix (See sec.~4 in~\cite{Antusch:2005gp}.). 
In the supersymmetric version, all contributions to the Majorana mass matrix 
receive the same quantum corrections, and hence zero mass eigenvalues remain zero.

\section{Screening, Grand Unification and flavor symmetry\label{sec:GUT}}

Let us discuss  the possible origin of the  
screening structure which results from the mass matrix (\ref{eq:matrix})
together with the condition~(\ref{eq:relation}). The texture of the 
matrix ~(\ref{eq:matrix})  with zero 1-1, 1-3, and 2-2 blocks can 
be obtained, {\it e.g.},  assigning the lepton numbers 
\begin{equation*} 
L(\nu) = L(S) = 1, ~~~L(N) =  -1, ~~~ L(\phi) = 0, ~~~ L(\sigma) = 0 
\end{equation*}
so that the lepton numbers of the blocks in the mass matrix (\ref{eq:matrix})
equal 
\begin{equation}
 \label{eq:matrixL}   
 L(\mathcal{M}) =  \left(
      \begin{array}{ccc}
        2 & 0 & 2\\
        0 & -2 & 0\\
        2 & 0 & 2\\
      \end{array}
    \right)\; .
  \end{equation}
Then the texture (\ref{eq:matrix}) appears if the lepton number is 
only broken by the Majorana mass terms of $S$. The lepton number can 
be broken explicitly or spontaneously by the VEV of the new scalar 
field $\rho$ which  has the lepton number $L(\rho) = - 2$ and couples 
with $S$ only: $S^T Y_S S \rho$. (The coupling of Majoron with the SM fields in negligible.)  
The  interaction  
$\nu^T  S \rho$ is forbidden by the gauge symmetry. The possible 
non-renormalized term 
\begin{equation*}
\frac{1}{M_{Pl}}  L  S \phi \rho
\end{equation*}
gives negligible effects due to small VEV $\braket{\phi}$. The coupling 
of $\rho$ with $\nu$ is also forbidden by the gauge symmetry. The term
$NN \rho$ is absent in the supersymmetric version due to holomorphy. 

In the non-supersymmetric version or if also the left 
superfield  $\rho^c$   exists,  an extended gauge symmetry can forbid the 
2-2 entry. Indeed, in left-right symmetric models $N$ enters 
the doublet of $SU(2)_R$ and the 2-2 block has gauge charge 
(1,3). The corresponding mass  term appears once the Higgs triplet 
$\Delta_R$ exists. 

The whole texture (\ref{eq:matrix}) can be a consequence of the gauge 
symmetry. Let us consider the $SU(2)_L\times SU(2)_R \times U(1)$ symmetry~\CiteLR.
The $[SU(2)_L, SU(2)_R]$ gauge properties of the mass matrix elements are  
\begin{equation}
 \label{eq:matrixG}   
 G(\mathcal{M}) = \left(
      \begin{array}{ccc}
        [3,1] & [2,2] & [2,1]\\
        ... & [1,3]  & [1,2]\\
        ... & ...    &  [1,1]\\
      \end{array}
    \right)\; .
  \end{equation}
The required matrix structure is  generated if the Higgs bi-doublet  
with the electroweak VEV, the RH doublet with GU-scale VEV and the 
singlet with $M_{Pl}$ scale VEV exist.  No particular lepton number 
prescription is needed. 

In the context of SO(10)~\CiteSOten, $\nu$ and $N$ belong to the 16-plet and 
$S$ is a singlet. The required texture can be generated by the 
following Yukawa interactions: 
\begin{equation}
\label{eq:so10}
Y_{\nu}~ {\bf 16} \times {\bf 16} \times {\bf 10}_H  + 
Y_N~  {\bf 16} \times {\bf 1} \times {\bf \overline{16}}_H
+  Y_S~ {\bf 1}\times {\bf 1} \times {\bf 1}_H , 
\end{equation}
where  ${\bf 10}_H$,  ${\bf \overline{16}}_H$, and ${\bf 1}_H$ are the 
Higgs multiplets. To generate matrix (\ref{eq:matrix})  
${\bf 10}_H$ should acquire the electroweak VEV, 
${\bf \overline{16}}_H$  -  the GU scale VEV in  $N$ ($SU_5$ singlet) 
direction and ${\bf 1}_H$ -  the Planck scale VEV. 

The interactions (\ref{eq:so10}) do not produce quark mixing,  and the Dirac 
masses of quarks and leptons are equal at the GUT scale. 
So, realistic model should contain some additional  sources  of the 
fermion masses which may, in general,   destroy screening. 
For instance, one can introduce $126$-plet of Higgses. Then 
VEVs of 126-plet in the  ``triplet'' directions 
will generate the 1-1 and 2-2 blocks and therefore  should be small
enough.  An alternative is the non-renormalizable interactions of the
type  ${\bf 16} \times {\bf 16} \times {\bf 16}_H \times {\bf 16}_H /M_{Pl}$. \\

Apparently, the interactions (\ref{eq:so10}) do not lead to relation (\ref{eq:relation}). 
The equality or proportionality of the  Yukawa couplings (\ref{eq:relation}) can 
appear due to 
\begin{itemize}
\item 
further unification  of $\nu$ and $S$;  
\item
certain flavor symmetry. 
\end{itemize} 
Let us comment on these two possibilities.  

1). The neutral leptons $\nu$, $N$ and $S$, can be embedded 
into a single representation $\irrep{27}$ of the gauge symmetry group
$E_6$~\CiteESix. 
Notice that in this case there are two additional neutral leptons in each 
generation: $S'$ and $S''$.  The screening structure - the matrix
(\ref{eq:matrix}) with the equality (\ref{eq:relation}) can be 
generated by the couplings 
\begin{equation}
  Y_{27}~ {\irrep{27}}\times {\irrep{27}}\times {\irrep{27}}_H +
  Y_{351_S}~ {\irrep{27}}\times {\irrep{27}}\times
  \left({\irrep{351_S}}\right)_H
  +
  Y_{351_A}~ {\irrep{27}}\times {\irrep{27}}\times \left({\irrep{351_A}}\right)_H\; ,
\end{equation}
where the 27-plet as well as the  symmetric and antisymmetric 351-plets of Higgses are introduced.  

In terms  of the maximal subgroup  
$\SU(3)_L\times\SU(3)_R\times\SU(3)_C\subset E_6$, the leptons transform as
$L\sim\left(\irrep{\overline{3}},\, \irrep{3},\, \irrep{1}\right)$.   
The $\left(\SU(3)\right)^3$ assignment of the neutral leptons is 
$$
\nu\sim L_3^{\dot{2}},\, ~N\sim L_2^{\dot{3}},\,~
S \sim  L_3^{\dot{3}},\, ~S'\sim L_1^{\dot{1}} ,\, ~S''\sim L_2^{\dot{2}}.
$$
The neutral components of Higgs multiplets $H$, $H_A$ and $H_S$ which can 
acquire VEVs  belong to 
\begin{align*}
  H \subset \left(\irrep{\overline{3}},\, \irrep{3},\, \irrep{1}\right) \subset &\;
  {\irrep{27}}_H \\
H_S \subset \left(\irrep{\overline{3}},\, \irrep{3},\, \irrep{1}\right)+\left(\irrep{6},\, \irrep{\overline{6}},\,
  \irrep{1}\right) \subset &\;\left({\irrep{351_S}}\right)_H\\
H_A \subset \left(\irrep{\overline{3}},\, \irrep{3},\, \irrep{1}\right)+\left(\irrep{\overline{3}},\, \irrep{\overline{6}},\, \irrep{1}\right)+\left(\irrep{6},\, \irrep{3},\, \irrep{1}\right) \subset &\;\left({\irrep{351_A}}\right)_H\; . 
\end{align*}
The Higgs multiplets $H$ and $H_A$ generate the Dirac structure\footnote{The
${\irrep{27}}_H$ and $\left({\irrep{351_A}}\right)_H$ can not generate 
Majorana mass terms because the corresponding Yukawa interactions 
 has to be antisymmetric in the $\SU(3)$ indices.} and the Majorana mass
terms are generated by $H_S$~\cite{Stech:2003sb}. 

Notice that it is not possible to get a  Dirac mass term of $S = 
L^{\dot{3}}_3$  with   $N = L^{\dot{3}}_2$, using  ${\irrep{27}}_H$ Higgs multiplet due to
antisymmetric (in SU(3) indices) interactions. 
The symmetric Higgs representation  $\left({\irrep{351_S}}\right)_H$ can generate  all 
mass terms of neutral leptons  required for screening.  
Indeed, the VEVs of $H_{\{\dot{2}\dot{3}\}}^{\{23\}}$ 
and $H_{\{\dot{2}\dot{3}\}}^{\{33\}}$ 
can be  of order of the electroweak scale and of the $\SU(2)_R$ breaking scale correspondingly. 
Also its  $H_{\{\dot{3}\dot{3}\}}^{\{33\}}$ component generates the Majorana 
mass of $S$. However,  with a single $\left({\irrep{351_S}}\right)_H$, the structure of matrix 
$M_S$ will be the same as the structure of the Dirac mass matrices.
One can introduce a second  $\left({\irrep{351_S}}\right)_H$ which produces the mass matrix
of $S$ of different structure.  

Another more promising possibility is 
to  use the antisymmetric $\left({\irrep{351_A}}\right)_H$ Higgs multiplet which can 
generate all necessary Dirac matrices.
It does not produce  the Majorana masses of $S$ which can be done using 
$\left({\irrep{351_S}}\right)_H$ so that the structure of $M_S$ is different from that of Dirac 
structures.

The following VEVs of the  $\left({\irrep{351_A}}\right)_H$ and  $\left({\irrep{351_S}}\right)_H$  
components~\footnote{The upper indices are $\SU(3)_L$ indices in the
  fundamental ($\irrep{3}$) representation and the lower
  ones belong to $\SU(3)_R$. The $\irrep{6}$ of $\SU(3)$ is represented by symmetric
  $3\times3$ matrices and described by 2 symmetrized indices. Dotted indices
  belong to the complex conjugate representation ($\irrep{\overline{3}}$). Flavor indices are
suppressed.} 
\begin{align*}
\braket{\left(H_A\right)_1^{\dot{1}}} \simeq&\; \Ord{\SU(2)_L\; \mathrm{breaking\;scale}}\\
\braket{\left(H_A\right)_1^{\{33\}}}\simeq&\;\Ord{\SU(2)_R\; \mathrm{breaking\;scale}}\\
\braket{\left(H_S\right)^{\{33\}}_{\{\dot{3}\dot{3}\}}}\simeq\braket{\left(H_A\right)_3^{\dot{3}}}\simeq&\;\Ord{\SU(3)_L\times\SU(3)_R\; \mathrm{breaking\;scale}}
\end{align*}
lead to the double seesaw.   
Indeed, in the basis  
$\left( \nu,\, N,\,S ,\, S',\, S''\right)$ they generate mass matrix
\begin{equation}
\label{eq:nLEsix}
  \left(\begin{array}{ccccc}
    0 & -Y_{351_A} \braket{\left(H_A\right)_1^{\dot{1}}} & 0& 0& 0 \\
    \dots & 0 & -Y_{351_A}\braket{\left(H_A\right)_1^{\{33\}}} & 0 &0 \\
    \dots & \dots & Y_{351_S}
    \braket{\left(H_S\right)^{\{33\}}_{\{\dot{3}\dot{3}\}}} &  Y_{351_A} \braket{\left(H_A\right)_1^{\dot{1}}}&  0\\
    \dots & \dots & \dots & 0 & Y_{351_A} \braket{\left(H_A\right)_3^{\dot{3}}}\\
    \dots & \dots & \dots & \dots & 0  \\
  \end{array} 
\right)\; 
\end{equation}
with the required structure for $\nu$,  $N$ and $S$. 
In addition to the mass spectrum of the double seesaw, there is one pseudo-Dirac
particle formed by $S'$ and $S''$ with 
a mass of the order of the $\SU(3)_L\times\SU(3)_R$ breaking scale.

Notice that interactions with ${\irrep{27}}_H$ Higgs multiplet can be used to 
generate sub-leading effects, correcting the masses of quarks and
producing some deviation from complete screening if needed. Furthermore, as the
VEVs of components contributing to the 1-3 and 2-2 element break $\SU(2)_L$ invariance,
they can only lead to entries of the order of the electroweak scale.

2). The equality of the Yukawa matrices (\ref{eq:relation}) can be a
consequence of certain flavor symmetry (See
reviews~\cite{King:2003jb,Altarelli:2004za} and references therein.).  It implies that  
the flavor charges of $\nu$ and $S$  are equal and the flavor symmetry  uniquely 
determines  the Yukawa couplings of these components with $N$. 
For this to happen, the flavor symmetry should be non-Abelian. 

In the Froggatt-Nielsen (F-N) approach~\cite{Froggatt:1978nt}  with U(1) flavor symmetry
(and in SO(10) context) we can write 
  \begin{equation}
\label{eq:fn}
\left(Z_{ab}  {\bf 16}_a \times {\bf 16}_b \times {\bf 10}_H  +  
Z_{ab}'{\bf 16}_a {\bf 1}_b \times {\bf \overline{16}}_H \right) 
\left(\frac{H_{FN}}{M_{FN}} \right)^{q_a + q_b}, 
\end{equation}
where $H_{FN}$ is the F-N scalar with $U(1)$ charge -1, 
$M_f$ is the scale where the operators (\ref{eq:fn}) are formed and 
 $q_a$ are the $U(1)$ charges of fermionic fields ${\bf 16}_a$. 
$Z_{ab}$ and  $Z_{ab}'$ are supposed to be couplings of the order 1, but  
for screening to work one needs to impose an additional proportionality condition
$Z_{ab} = c Z_{ab}'$ ($c = const$). \\

Let us comment on a completely different possibility. If  $S$ belongs to the
16-plet  and $N$ is the singlet of SO(10), the required equality
of the Yukawa couplings (\ref{eq:relation}) is automatically reproduced.  
The screening structure can be generated by interactions  
\begin{equation}
\label{eq:so10b}
Y {\bf 16} \times {\bf 1} \times {\bf \overline{16}}_H
+  Y_S~ {\bf 16}\times {\bf 16} \times {\bf \overline{126}}_H +  
Y_q {\bf 16} \times {\bf 16} \times {\bf 10}_H \; ,
\end{equation}
if ${\bf \overline{16}}_H$ has the electroweak VEV in the ``$\nu_L$'' direction and 
the GUT scale VEV in the ``$\nu_R$'' 
direction, 
and ${\bf \overline{126}}_H$ has the Planck scale VEV in the direction of the
SU(5) singlet. The last term in (\ref{eq:so10b}) gives the Dirac 
masses of quarks and leptons and also the Dirac mass term for $\nu$ 
and $S$. The mass matrix generated by (\ref{eq:so10b})  equals 
\begin{equation}
 \label{eq:matrixD}   
 \mathcal{M} =  \left(
      \begin{array}{ccc}
        \sim 0 & Y \braket{{\bf \overline{16}}_H} &  Y_q \braket{{\bf 10}_H}  \\
        Y^T \braket{{\bf \overline{16}}_H}^T & \sim 0 & Y^T \braket{{\bf \overline{16}}_H}^T\\
        Y_q \braket{{\bf 10}_H}  & Y \braket{{\bf \overline{126}}_H} & Y_S \braket{{\bf \overline{126}}_H}\\
      \end{array}
    \right)\; .
\end{equation}
The 126-plet can also contribute to the 1-1 and 1-3 blocks. 
However, now $Y_\nu$ and $Y_N$  are not related to 
the Dirac matrices of quarks, and the problem of screening does not
exist from the beginning. The usual seesaw contribution due to 
Dirac matrices which are related  to the quark mass matrices 
$Y_q$ is suppressed by large (Planck) scale.  
That is,  here we deal with suppression and not screening of the Dirac structures. 
Also in this case it would be more
natural to identify the RH neutrino with $S$.

\section{Conclusions\label{sec:conclusions}}

1). We studied  screening of the Dirac  flavor
structure  in the neutrino masses and mixings. 
Screening is realized  in the context of double seesaw mechanism if the Dirac type  
Yukawa couplings of both seesaw steps are proportional to each other. 
The flavor structure of the light neutrino mass matrix is 
determined (in lowest order) by the  
flavor structure of  matrix $M_S$  - the Majorana mass matrix of new singlets at the Planck scale, 
while the Dirac flavor structure is completely screened. 
The scale of the neutrino masses is correctly determined by the 
scales involved in the double seesaw as $M_{Pl} v_{EW}^2/M_{GU}^2$.\\

\noindent
2). The Planck-scale Majorana mass matrix $M_S$ violates lepton number, while at the
same time it may have rich flavor symmetries. Screening translates
these symmetries of $M_S$ directly into symmetries of $m_\nu$.  
The structure of $M_S$ can therefore be the origin of specific 
``neutrino symmetries'' which do not show up in the quark sector.  
In particular, $M_S$ can be degenerate 
or quasi-degenerate leading (after the inclusion of radiative
corrections) to a quasi-degenerate spectrum of light neutrinos. 
$M_S$, and consequently $m_\nu$, can also have certain 
symmetries which result in maximal 2-3 mixing and small 1-3 mixing. 
A very interesting possibility is that $M_S$ may have the  
bi-maximal structure which explains the quark-lepton complementarity. 
Screening allows therefore to reconcile the quark-lepton symmetry 
expressed as (approximate) equality of the Dirac mass matrices of 
quarks and leptons with the drastically different patterns of 
quark and lepton mixings. 
Thus the  screening mechanism offers interesting possibilities in GUT 
model building. \\

\noindent
3). The screening mechanism involves the cancellation of mass 
matrices which arise upon symmetry breaking at very different 
energy scales, namely the electroweak scale and the GUT scale. 
These matrices have  different gauge properties leading to differences in the radiative corrections which
can destroy screening. We studied stability of screening with 
respect to renormalization group effects. 

It has been  found that screening 
is stable in the MSSM and the external renormalization of the 
effective d=5 operator can be small. This means that the structure 
of the light neutrino mass matrix is still mostly determined by 
$M_S$, with small radiative corrections after renormalization 
group running is included. 

In contrast, in the case of the SM, 
due to the d=5 operator corrections, the screening is unstable for 
certain structures of $M_S$. The form of the light neutrino mass
matrix $m_\nu$ can differ  strongly from that of $M_S$.
This can, in fact, be used to explain some features of the 
light neutrinos via renormalization group effects. \\

\noindent
4). Finally, the structure of the mass matrix (\ref{eq:matrix}) which leads to screening can 
be obtained using the lepton number or/and gauge symmetry. 
We outlined how screening could be realized in GUTs. The equality 
of the Dirac type Yukawa coupling matrices of $\nu$ and $N$ and 
of $S$ and $N$ can be a consequence of  certain (horizontal) 
flavor symmetry or further unification of the fields. 
The latter can be realized, {\it e.g.}, within the framework of $E_6$ gauge models.

\section*{Acknowledgment}

The authors are grateful to B. Stech for valuable comments.
The work of A.Yu.S. was supported by  Alexander von Humboldt Foundation 
(the Humboldt Research Award). This work was also supported by the 
``Deutsche Forschungsgemeinschaft'' in the 
``Sonderforschungsbereich 375-95 f\"ur Astro-Teilchenphysik'' 
and under project number RO-2516/3-1.

\appendix
\renewcommand{\theequation}{\thesection.\arabic{equation}}
\setcounter{equation}{0}
\section{Appendix\label{sec:renfactors}}

We use GUT charge normalization for the hypercharge, {\it i.e.}
$\frac{3}{5}\left(g_1^\mathrm{GUT}\right)^2=\left(g_1^\mathrm{SM}\right)^2$. The
factors $Z$ describing the LL approximation are obtained from the counterterms
in Ref.~\cite{Antusch:2002rr}.

\subsection{SM\label{sec:appone}}
In the SM extended by RH neutrinos, the wave function renormalization of the RH neutrinos is given by
\begin{equation}
\EFT{n}{Z_N}=\exp\left(\frac{1}{16\pi^2}\EFT{n}{Y_\nu^\dagger} \EFT{n}{Y_\nu}\ln\frac{M_n}{M_{n+1}}\right)\\
\end{equation}
and collecting the contributions from the renormalization of the
left-handed doublets
\begin{equation*}
  \beta^L_{Y_\nu}=\frac{1}{32\pi^2}\left(\EFT{n}{Y_\nu} \EFT{n}{Y_\nu^\dagger} +
  Y_eY_e^\dagger\right)\label{eq:ZL}\; ,
\end{equation*}
the Higgs doublet
\begin{equation*}
\beta_{Y_\nu}^\phi=\frac{1}{32\pi^2}\left(2 \tr\left(\EFT{n}{Y_\nu} \EFT{n}{Y_\nu^\dagger}
+
Y_eY_e^\dagger+3Y_uY_u^\dagger+3Y_dY_d^\dagger\right)-\frac{9}{10}g_1^2-\frac{9}{2}g_2^2\right)\; ,
\end{equation*}
and the vertex correction to $Y_\nu$
\begin{equation*}
\beta_{Y_\nu}^{Y_\nu}=-\frac{1}{8\pi^2}Y_eY_e^\dagger\label{eq:ZYnu}
\end{equation*}
the external renormalization in the effective theory with $n$ RH neutrinos yields
\begin{multline}
  \EFT{n}{Z_\mathrm{ext}}=\exp\left(\frac{1}{32\pi^2}\left(\EFT{n}{Y_\nu}
      \EFT{n}{Y_\nu^\dagger} -3 Y_eY_e^\dagger+2 \tr\left(\EFT{n}{Y_\nu}
      \EFT{n}{Y_\nu^\dagger}+Y_eY_e^\dagger+3Y_uY_u^\dagger+3Y_dY_d^\dagger\right)\right.\right.\\
  \left.\left.-\frac{9}{10}g_1^2-\frac{9}{2}g_2^2\right)\ln\frac{M_n}{M_{n+1}}\right)\; .
\end{multline}
Neglecting the thresholds in the charged lepton sector and the quark sector, the
expression for the external renormalization factor $Z_\mathrm{ext}^\mathrm{SM}$
describing the total external renormalization can be further approximated to
\begin{multline}
Z_\mathrm{ext}^\mathrm{SM} = 
\exp\left(\frac{1}{32\pi^2}\sum_{n=0}^3\left[\EFT{n}{Y_\nu}\EFT{n}{Y_\nu^\dagger} + 
2\tr\left(\EFT{n}{Y_\nu}  \EFT{n}{Y_\nu^\dagger}\right) \right]\ln\frac{M_n}{M_{n+1}}\right.\\
\left.+\frac{1}{32\pi^2}\left[-3 Y_eY_e^\dagger+
2\tr\left(Y_eY_e^\dagger+3Y_uY_u^\dagger+3Y_dY_d^\dagger\right) - 
\frac{9}{10}g_1^2-\frac{9}{2}g_2^2\right]\ln\frac{\braket{\phi}}{\Lambda}\right). 
\end{multline}
Here for uniformity of presentation we have denoted 
\begin{equation*}
M_0 \equiv \braket{\phi}, ~~~~  M_4 \equiv \Lambda\; . 
\end{equation*}
The renormalization effect due to the additional vertex corrections to the
d=5 operator is given by 
\begin{equation}
\EFT{n}{Z_{\kappa}}=\exp\left(\frac{1}{16\pi^2}\left(\lambda+\frac{9}{10}g_1^2+\frac{3}{2}g_2^2\right)\ln\frac{M_n}{M_{n+1}}\right) \; .
\end{equation}
The mass of the right-handed neutrinos receives only corrections from the wave
function renormalization to arbitrary loop order.
%

\subsection{MSSM\label{sec:apptwo}}

In the MSSM extended by RH neutrinos, there are no vertex corrections due to the non-renormalization
theorem and the wave function renormalization yields
\begin{align}
\EFT{n}{Z_L}=&\exp\left(\frac{1}{32\pi^2}\left(2Y_eY_e^\dagger+2\EFT{n}{Y_\nu}
  \EFT{n}{Y_\nu^\dagger}-\frac{3}{5}g_1^2-3 g_2^2 \right)\ln\frac{M_n}{M_{n+1}}\right)\\
\EFT{n}{Z_N}=&\exp\left(\frac{1}{8\pi^2}\EFT{n}{Y_\nu^\dagger} \EFT{n}{Y_\nu}\ln\frac{M_n}{M_{n+1}}\right)\\
\EFT{n}{Z_\phi}=&\exp\left(\frac{1}{32\pi^2}\left(\tr\left(6Y_uY_u^\dagger+2\EFT{n}{Y_\nu} \EFT{n}{Y_\nu^\dagger}\right)-
\frac{3}{5}g_1^2-3g_2^2\right)\ln\frac{M_n}{M_{n+1}}\right)\; .
\end{align}
The external renormalization factor $Z_\mathrm{ext}^\mathrm{MSSM}$ is given
by the product of the wave function renormalization of the left-handed doublet
with the Higgs doublet 
\begin{equation}
  \EFT{n}{Z_\mathrm{ext}}=\EFT{n}{Z_L}\EFT{n}{Z_\phi}
\end{equation}
because the two wave function renormalization factors commute.
As the neutrino Yukawa couplings only change at the thresholds (up to 1 loop
order), the external renormalization factor can be further approximated by
\begin{multline}
Z_\mathrm{ext}^\mathrm{MSSM}=\exp\left(\frac{1}{16\pi^2}\sum_{n=0}^3\left(\EFT{n}{Y_\nu}
  \EFT{n}{Y_\nu^\dagger} + 
\tr\left(\EFT{n}{Y_\nu}\EFT{n}{Y_\nu^\dagger}\right)\right)\ln\frac{M_n}{M_{n+1}}\right.\\
\left. + \frac{1}{16\pi^2}\left(Y_eY_e^\dagger-\frac{3}{5}g_1^2-3 g_2^2 
+3\tr\left(Y_uY_u^\dagger\right)\right)\ln\frac{\braket{\phi}}{\Lambda}\right)\; .
\end{multline}

\end{fmffile}

\clearpage


\providecommand{\bysame}{\leavevmode\hbox to3em{\hrulefill}\thinspace}

\end{document}